# Experimental Study on the Aerodynamic Sealing of Air Curtains

**João Carlos Viegas [1,\*], Fernando Oliveira [2] and Daniel Aelenei [2]**

[1] National Laboratory of Civil Engineering, Av. do Brasil 101, 1700-066 Lisbon, Portugal
[2] Faculty of Sciences and Technology, Universidade Nova de Lisboa, 2829-516 Caparica, Portugal; fm.oliveira@campus.fct.unl.pt (F.O.); aelenei@fct.unl.pt (D.A.)
\* Correspondence: jviegas@lnec.pt; Tel.: +351-914-741-030

**Abstract:** Controlling the air quality is of the utmost importance in today's buildings. Vertical air curtains are often used to separate two different climatic zones with a view to reduce heat transfer. In fact, this research work proposes an air curtain aimed to ensure a proper separation between two zones, a clean one and a contaminated one. The methodology of this research includes: (i) small-scale tests on water models to ensure that the contamination does not pass through the air curtain, and (ii) an analytical development integrating the main physical characteristics of plane jets. In the solution developed, the airflow is extracted from the contaminated compartment to reduce the curtain airflow rejected to the exterior of the compartment. In this research work, it was possible to determine the minimum exhaust flow necessary to ensure the aerodynamic sealing of the air curtain. This article addresses the methodology used to perform the small-scale water tests and the corresponding results.

## 1. Introduction

Air curtains consist of plane jets that are frequently used to separate different environmental zones. Air curtains are basically designed to reduce or control both heat and mass transfers, as well as to reduce the spreading of airborne contaminants between two zones. Their application is particularly useful when the physical barriers are not viable, for several reasons. In this framework, air curtains have been employed for several purposes, such as: HVAC (Heating, Ventilation and Air Conditioning) [1–3], smoke control in passageways [4,5], airborne pollutant, and biological control [6–8].

The plane jet induces a shear stress between the airflow and the stagnant ambient air, which promotes air entrainment. In the jets with significant flow rates, the shear stress induces the development of turbulent structures with several length scales. The turbulent structures promote air mass entrainment and mixing from both sides into the jet ('contaminated' and 'non-contaminated' compartments). Furthermore, the rejection of the jet flow to the clean compartment corresponds to the loss of air curtain tightness (loss of pollutant containment).

The usual application of this technology to protect the non-contaminated compartment from airborne contaminants (microorganisms, bacteria, fungi, and particles), in an almost isothermal condition, uses low jet velocities for obtaining an approximately laminar flow. The extraction at the contaminated side or at the tip of the jet is necessary in order to avoid the dispersion of the jet flow into the non-contaminated compartment (Figure 1a). Reversely, this objective can also be attained if clean air is supplied to the non-contaminated compartment (Figure 1b). Treating the 'clean' air before supplying it to the non-contaminated compartment, or treating the 'dirty' air removed from the contaminated compartment before releasing it into the environment are expensive procedures. Therefore, it becomes essential to find a technical solution for creating a proper zone separation that will be able to minimize flow rate requirements.

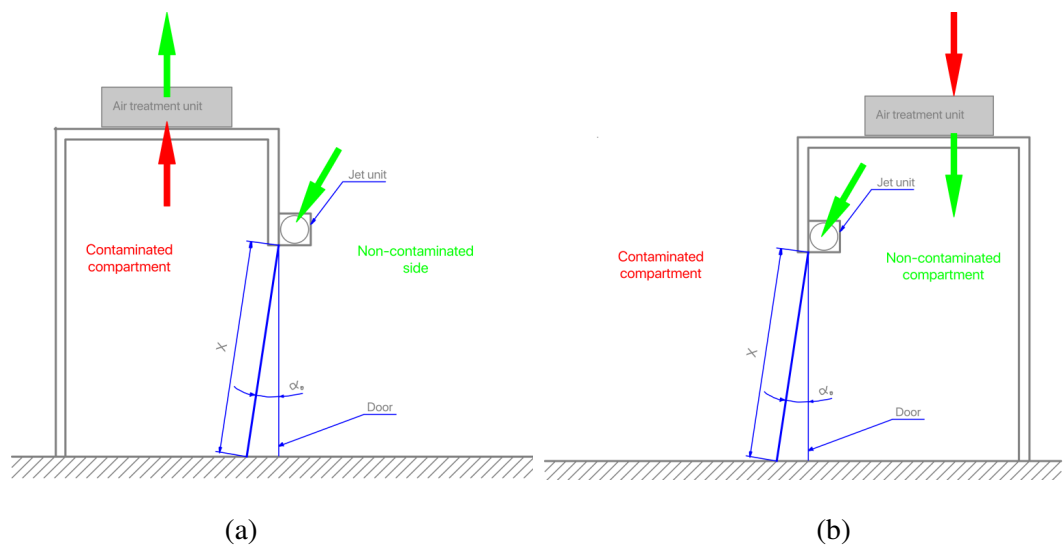

**Figure 1.** Strategies for the containment of a contaminant within a compartment (**a**) and for the protection of a clean room (**b**).

In the context of the Nanoguard2ar project, the basic requirement is to reduce, as far as possible, the exhaust flow rate from the contaminated compartment (or the supply flow rate to the non-contaminated compartment), in order to reduce the air treatment costs. Although, for an undisturbed environment (e.g., isothermal flow), the best solution is just to provide a very small uniform horizontal air stream through the opening, in practice, there are disturbances (e.g., temperature differences, wind, people crossing the opening) that require the use of an air curtain to improve the aerodynamic sealing of the opening. For this purpose, the most effective air curtain is the one that is able to reduce the airborne contaminants, minimizing the exhaust (or supply) flow rate, when exposed to disturbances.

Hayes and Stoecker [9–11] studied the aerodynamic sealing performance of vertical air curtains under isothermal and non-isothermal conditions. Hayes [8] developed the deflection modulus ($D_m$), which is indicative of the deflection of the air curtain jet and expresses the ratio of the outlet momentum to the transverse forces on the jet [12]. They defined [10] the operation condition where the air curtain jet reaches the floor as the 'optimum condition' and other operation conditions where the air curtain jet does not reach the floor as 'break-through condition', which can be further divided into the 'inflow breakthrough condition', where the air curtain flow is curved inward and does not reach the floor, and the 'outflow breakthrough condition', where the air curtain flow is curved outwards and does not reach the floor [13]. In our research, the ideal condition corresponds to the limit between the optimum condition and the inflow breakthrough condition, when having Figure 1 as reference (Figure 1a), because in this condition, the contaminant mixed up in the jet flow is not released to the outside of the compartment.

In the air curtain applications at door height openings, the flow is characterized, in practice, by a transition regimen. The overall jet flow rate is reduced if the jet velocity is decreased; and hence, the use of lower jet velocities is also a strategy that can be adopted to reduce the jet flow rate rejected to the 'non-contaminated' compartment. Nevertheless, the vertical downward air curtains must be properly designed so as to avoid splitting the jet flow at the floor impingement zone, and consequently losing the air curtain aerodynamic sealing. However, with low air curtain speeds, when people walk through the door, the air curtain reconstitution time is longer, which could increment pollutant leakage through the air curtain. Using air curtains in a transition regimen (rather than in a laminar regimen) can minimize this problem.

This study is developed within the framework of the Nanoguard2ar project (European Union's Horizon 2020 research and innovation programme under the Marie Sklodowska-Curie grant agreement N 690968). Its main goals are to develop, design, test, validate, and demonstrate an innovative nanomaterial-based 'microbial free' engineering solution, to ensure indoor air quality in buildings. To achieve this, an advanced oxidation process will be employed to clean the air extracted from the contaminated compartment (or to clean the external air before supplying it to the non-contaminated compartment), in conjunction with the use of air curtains, ensuring a proper separation between the spaces that are to be kept without cross contamination. Therefore, the purpose of this study is to define the plane jet characteristics that enable the best aerodynamic sealing of the air curtain to be achieved, by ensuring low exhaust flow rates from the contaminated compartment (or supply flow rates to the non-contaminated compartment), in accordance with the air treatment procedures adopted. This research encompasses three phases, namely: (i) small-scale experiments using water as the working fluid; (ii) computational fluid dynamics (CFD) simulations in order to verify if the small-scale test results are applicable to full size air curtains; and (iii) full size air curtain experiments. This paper describes the research carried out in the first phase.

There are several applications of this concept that aim to avoid contaminant spreading, such as: an operating room [7,14–19], a tobacco smoke control [20,21], the protection of art works in museums and of cultural heritage [22,23], and open refrigerated display cabinets [24,25]. Several studies about air curtain efficiency have been conducted on these applications. However, there is no systematic approach to the contaminant aerodynamic sealing in the transition between the optimum condition and the 'inflow breakthrough condition'. The studies on the application of air curtains have been based on experiments [13,26–29], on CFD models [23,29,30], and on semi-analytical models [12,31].

Rydock et al. [27] presented an experimental study about the efficiency of air curtains in a dedicated-smoking restaurant area. The air curtain was installed at a 0.8 m height and the slot widths corresponded to 10 mm, 5 mm, and 3 mm, respectively, with varying mounting angles (+15, 0, and −15), and varying supply and extraction airflow rates, as well as several non-smoking sections with different air supply configurations. The maximum air curtain efficiency was achieved using the 100 m$^3$/h per meter of air curtain and for a slot width of 5 mm. The results seem to indicate the feasibility of the use of air curtains to achieve improved air quality. However, a smoke free environment cannot be completely attained in a section of a single room. In this study, the air curtain flow was fed by the contaminant air from the smoking zone. The authors of this work concluded that to achieve an aerodynamic sealing for the tobacco smoke, the air curtain flow must be fed by clean air.

Shih et al. [17] presented a numerical study on ethanol spreading in clean rooms. For the purpose of the study, the authors analysed the influence of air curtain parameters, such as jet velocity, angle, and installation height. The minimum velocity studied was 3 m/s. The maximum efficiency was achieved for a jet speed of 5 m/s and an angle of 15°. In this study, it was observed that the loss of air curtain sealing occurred essentially close to the bottom, where a flow separation takes place.

Santoli, Cumo, and Mariotti [23] presented a numerical and experimental study on a cultural heritage building, using an air jet with speeds between 4 and 5 m/s, to create a physical barrier to the flow. They concluded that the ventilation system has efficiencies of 70–75%. However, the parameters that allow for maximizing the efficiency of the air curtain were not analysed in this study.

In operating rooms, the critical operating area is often isolated from the external contaminants by the installation of laminar diffusers, in conjunction with the use of air curtains [16]. Cook and Int-Hout [18] pointed out that another way to guarantee the asepsis level in the operating room is to ensure a hierarchical pressure in this space, between 10 to 15% of the volume of the room. According to the authors, the difficulty in guaranteeing a high asepsis level is due to the difficulty in achieving a laminar flow in the panels. This difficulty is associated with the undesirable speeds attained by the air entrainment and with the pressure gradient resulting from temperature differences, which eventually increase the turbulence intensity in the room. However, these studies are not conclusive with regards to the speed that should be used to achieve a certain level of asepsis. They only refer to the fact that low speeds are more efficient [16].

Zhai and Osborne [14] presented a numerical and experimental study, in which they concluded that there is no correlation between the laminar flow diffuser, the air curtain flow rate, and the concentration of contaminant. It was observed that it was preferable to use one-way diffusers rather than air curtains with high jet velocities. However, this did not demonstrate the existing relation between the inflow and outflow rates.

Goubran et al. [13] present an experimental study to verify and further investigate the flow characteristics of the building entrances equipped with air curtains. In this study, the working fluid is the air and the nozzle thickness was 0.0635 m. Two different supply velocities were selected, namely: (1) the maximum supply speed of the unit is 13.75 m/s; and (2) a lower speed 9.1 m/s. A small-scale chamber (1/3 scale of the real case) of 2.44 m × 2.44 m × 1.3 m (L × W × H) was used for the measurements of the infiltration/exfiltration and pressure differentials, which were then used for developing the empirical model across the operating air curtain. The flow and pressure measurements confirmed that, for the tested pressure difference range, air curtains can significantly reduce the infiltration. Additionally, the experimental results indicated that the higher the air curtain supply speed, the better the air curtain performs (i.e., it is able to continue to operate in the optimum condition at higher pressure differences). Following this work, Qi et al. [29] presented a parametric study of air curtain performance based on reduced-scale experiments and full-scale numerical simulations. Using the same experimental setup of previous research [13], they found that increasing the air curtain supply angle improves the air curtain performance when it is operated under the optimum condition and inflow breakthrough condition, but creates excessive exfiltration under the 'outflow breakthrough condition'. Increasing the supply speed of the air curtain generally improves the air curtain performance, whereas this improvement deteriorates with the increase of the supply angle under the outflow breakthrough condition.

These works clearly stress the relevance of the pressure difference between inside and outside the compartment, which is due to several effects, such as the buoyancy effect, the head loss of the flow through the opening, and many other factors including the wind effect and indoor systems operation; however, in the absence of the buoyancy effect, this pressure difference through the opening protected by the air curtain is related with the head loss of the flow through the opening protected by the air curtain. As the head loss (that is expressed in these works by a discharge coefficient) is dependent on the characteristics of the opening [32], in our research, we considered that the average velocity of the flow across the door ($\overline{u_a}$) is the variable of reference.

In the mentioned works, no clear correlation was established between the inlet flow rate (flow entering the door opening) and the velocity at the jet nozzle, which assures both the maximum air curtain effectiveness, and the level of asepsis achieved under isothermal conditions or low temperature differences. As there is no concern in reducing the exhaust flow from the contaminated compartment (or the supply flow to the non-contaminated compartment), in order to increase the effectiveness of the curtain and to reduce the air cleaning costs, both the air velocities and the corresponding Reynolds numbers in these works are quite high when compared with those reported in this research work. Therefore, to the best knowledge of the authors of this work, this study constitutes an added value to research in the field.

For the purposes of this work, it was considered as relevant to evaluate the balance between the average speed at the door opening and the jet characteristics, with a view to increase the air curtain effectiveness and thus enable its integration into the Nanoguard2ar system. Minimal requirements of average speed at the door are obtained for very low jet velocity. Therefore, very low jet speeds will also be used up to a maximum of 1.70 m/s. In previous research [4], the extreme condition of the curtain exposed to a high buoyancy action was studied; in our research presented here, the opposed extreme condition of non-buoyant flow is studied.

A simplified analytical model is presented in the next sections, with a view to discuss the variables that can influence the performance of the jet planes under a turbulent transition regimen. Two different conditions are analysed, namely the simplified balance of the jet momentum (which demonstrates a strong influence of the jet angle and is relevant for smaller Reynolds Number values—146, 367 and 687) and the comparison between the flow rate entrained by the jet from the non-contaminated zone, and the flow rate across the opening required to avoid the loss of aerodynamic sealing.

## 2. Methods

### 2.1. Analytical Model

Figure 2 presents a cross section of the plane jet system. The flow rate balance is presented in Equation (1), where $\dot{M}_1$, $\dot{M}_2$, and $\dot{M}_3$ are the jet flow rate at the floor impingement zone, the flow rate rejected to the contaminated side, and the flow rate rejected to the non-contaminated side, respectively.

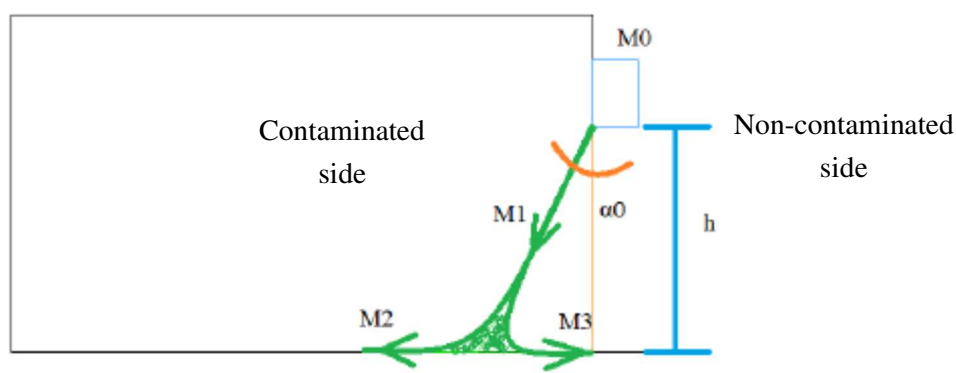

**Figure 2.** Cross section of the plane jet system.

In this derivation, the following assumptions are applicable:

- the momentum is conserved in the non-disturbed jet;
- the speed of the jet was assumed as uniform at the jet cross section; and
- the momentum balance is expressed by the product of the flow rate by the average speed (actually, this is not true and is only considered here to analyze the variables influencing the process; empirical correcting values will be afterwards deduced).

The momentum balance requires the definition of the quantities of Equations (2) to (5), as follows:

$$\dot{M}_1 = \dot{M}_2 + \dot{M}_3 \tag{1}$$

$$\dot{M}_0 = b_0 \times L \times \overline{u_0} \tag{2}$$

$$J_1 = b_0 \times L \times \overline{u_0}^2 \tag{3}$$

$$J_2 = \dot{M}_2 \times \overline{u_1} \tag{4}$$

$$J_3 = \dot{M}_3 \times \overline{u_1} \tag{5}$$

The average jet velocity at the floor impingement zone was also considered to be equal to the average velocity of the flow rate rejected to the non-contaminated side and to the average velocity of the flow rate rejected to the contaminate side, as presented in Equation (6):

$$\overline{u_1} = \overline{u_2} = \overline{u_3} \tag{6}$$

The momentum balance of the jet at the floor impingement zone is expressed by Equation (7), as follows:

$$J_2 = J_3 + J_1 \times \sin \alpha_0 \Leftrightarrow \dot{M}_2 \times \overline{u_1} = \dot{M}_3 \times \overline{u_1} + \dot{M}_1 \times \overline{u_1} \times \sin \alpha_0 \tag{7}$$

From Equation (7), it will be possible to estimate the split of the jet flow rate, in accordance with Equations (8) to (10), as follows:

$$\dot{M}_2 = \dot{M}_3 \times \frac{1 + \sin \alpha_0}{1 - \sin \alpha_0} \tag{8}$$

$$\dot{M}_2 = \dot{M}_1 \times \frac{1 + \sin \alpha_0}{2} \tag{9}$$

$$\dot{M}_3 = \dot{M}_1 \times \frac{1 - \sin \alpha_0}{2} \tag{10}$$

The momentum of the flow rate rejected to the non-contaminated side can be expressed by Equation (11), as follows:

$$J_3 = J_1 \times \frac{1 - \sin \alpha_0}{2} \Leftrightarrow J_3 = b_0 \times L \times \overline{u_0}^2 \times \frac{1 - \sin \alpha_0}{2} \tag{11}$$

It can be assumed that the momentum of the flow through the door $J_a$ (corresponding to the exhaust mass flow rate of the 'contaminated' compartment, Equation [12]) will be, at least, higher or equal to the momentum of the flow rate rejected to the 'non-contaminated' side. The limit condition is presented in Equation (13).

$$J_a = w \times h \times \overline{u_a}^2 \tag{12}$$

$$J_a = J_3 \Leftrightarrow w \times h \times \overline{u_a}^2 = b_0 \times L \times \overline{u_0}^2 \times \frac{1 - \sin \alpha_0}{2} \tag{13}$$

Assuming that w = L, the relation of Equation (14) can be deduced from Equation (13), as follows:

$$\overline{u_a} = \overline{u_0} \times \left[ \frac{b_0}{h} \times \frac{1 - \sin \alpha_0}{2} \right]^{0,5} \tag{14}$$

This equation will be used as reference in the analysis of the experimental results.

It was also assumed as relevant to compare another threshold condition with the flow rate through the door. In addition to the nozzle jet flow rate supplied from the non-contaminated compartment, there is also a part of the jet flow rate that is entrained from the non-contaminated compartment. Consequently, it can be assumed that the flow rate crossing the door will be, at least, equal to the flow rate entrained by the jet from the non-contaminated compartment. The jet flow rate is given by Equation (15) [33]:

$$Q_{jet} = 0.44 \times \left( \frac{2 \times x}{b_0} \right)^{\frac{1}{2}} \times Q_0 \tag{15}$$

Considering the values obtained from Equations (16) and (17), the threshold condition previously referred to is given by Equation (18), as follows:

$$Q_0 = b_0 \times L \times \overline{u_0} \tag{16}$$

$$x = \frac{h}{\cos(\alpha_0)} \tag{17}$$

$$\overline{u_a} = \left[ 0.22 \times \left( \frac{2 \times h}{b_0 \times \cos \alpha_0} \right)^{\frac{1}{2}} + 0.5 \right] \times \overline{u_0} \times \frac{b_0}{h} \tag{18}$$

Although Equations (14) and (18) do not present explicitly the pressure difference between the interior of the compartment and the exterior, such correlation can be obtained from these equations if the discharge coefficient corresponding to the opening protected by the plane jet is introduced. Therefore, the influence of the pressure difference is indirectly considered in this equation through $\overline{u_a}$.

Although in a previous work [4] the deflection modulus (defined by Hayes [9]) was included in the physical analysis, in this research, it is not necessary to consider this non-dimensional variable, because no buoyancy is acting in these experiments.

*2.2. Experimental Water Modelling*

In this research, tests were carried out on a small-scale model (1/20) installed in a water tank, with a view to simulate the full-size prototype (air curtain). Water was adopted as the testing fluid in order to maintain, in the model, the same range of Reynolds number as in the prototype. The definition of the Reynolds number currently used for plane jets, Re = $(\overline{u_0} b_0)/\nu$, was considered [33]. Although axisymmetric jets remain laminar up to Re = 1000 and became fully turbulent for Re > 3000, the plane jets from a long and narrow slot may present appreciable turbulent instabilities beyond Re as low as 30 [33]. Therefore, these small-scale tests, with Re ranging from 147 to 2125, will be carried out in the transition regimen. The results may not be easily extrapolated for full size air curtains unless complementary work will be done. For this reason, this research project encompasses further phases of full-scale CFD simulation of air curtains and final full-scale testing. The model was placed in a tank measuring 1.20 m × 0.49 m × 0.50 m, and the model compartment had the following dimensions 0.40 m × 0.25 m × 0.26 m, and comprised an opening that simulated the door with the dimensions of 0.125 m × 0.125 m. The jet was located at the soffit of the opening and had a width of 0.1456 m and thicknesses of 1.25 mm, 2.50 mm, 3.75 mm, and 5.00 mm, respectively. The jet was connected to a pump with a flowmeter that was used to measure the flow in the jet nozzle. In the compartment, on the opposite end of the door, there were two openings, one for the contaminant supply (fed by gravity and consisting of water with coloring), with a diameter of 0.01 m, and another for the exhaust flow (with a pump), with a diameter of 0.05 m. The contaminant level was kept constant so as to ensure that its flow rate also remained constant (Figure 3). In the photo of Figure 3, the red bottles were used to avoid the floating of the model. The model is upside down because, when used for studying non-isothermal flows, it is less expensive to simulate the hot fluid with saltwater, which is denser [4]. This position does not influence the isothermal results.

As previously mentioned, the jet complied with a set of geometric parameters, among which only the jet thickness and the jet angle (four angles used) were changed from test to test. In the tests from 89 to 91, the same 15° angle was invariably used and different nozzle thicknesses were adopted, while tests 55 and 56 were performed with an inactive jet. Table 1 indicates the geometric characteristics of the nozzle and the opening in the model, which is protected by the jet.

As this test is representative of the isothermal conditions, before initiating it, care was taken to ensure a constant water supply temperature, which was equal to the temperature of both the tank and the contaminant. Before every test, the jet angle and thickness were adjusted and the jet nozzle flow rate was set to the desired value, and was measured. During the test, the contaminant was released into the compartment and the exhaust flow rate was increased until no contaminant transport through the opening was visible. In the end of every test, the extraction flowrate from the compartment, corresponding to the contaminant sealing, was measured and was used to calculate the average velocity through the opening protected by the plane jet, considering the mass flow balance. No pressure difference measurements were carried out because its effect, the average velocity through the opening, was used in this assessment.

The standard uncertainty [34] estimated for the nozzle area ranged from 3% to 13%. The standard uncertainty estimated for the nozzle flow rate measurement was less than 0.5%. The standard uncertainty estimated for the nozzle velocity calculation ranged from 4% to 13%. The standard uncertainty estimated for the exhaust flow rate measurement corresponded to 2.5%.

In the test plan adopted, each selected variable was changed successively. Therefore, the tests started with a higher Reynolds number (with the values Re = 2125, Re = 1710, and Re = 1224) and a jet thickness of 1.25 mm, after which the jet velocity was reduced to obtain lower Reynolds number values of Re = 687, Re = 367, and Re = 147, respectively. After these tests, the variation in the jet thickness was studied for the following values: 2.50 mm, 3.75 mm, and 5.00 mm, respectively. In the study of the jet thickness, the Reynolds number Re = 367 was maintained, which is why the jet speed was reduced. Another additional test was carried out with the inactive jet under isothermal conditions and using the same method as the one previously described.

During the tests, the contaminant transport through the door was assessed by naked eye observation. Reference must be made of the fact that the human eye is not an optimal instrument, and, as such, it was necessary to establish a criterion to determine whether the exhaust flow was in a minimum limit state, so that no contaminant was leaking through the curtain. Therefore, the following cases were considered (Figures 4–6 present a detail of the test setup where is possible to see just the cross section of the plane jet and of the opening; these pictures were rotated in order so that the jet is presented in a downward position; the inside water is coloured by the turbulent mixture with the jet coloured water): oversized exhaust flow rate (Figure 4), limit state condition (Figure 5), and insufficient exhaust flow rate (Figure 6). In the first case, a triangle of discoloration in the water was observed in the compartment, which showed that the inflow at the door was too high; in the third case, some eddies were observed, which were leaking through the door (meaning that the hydrodynamic sealing of the curtain was lost); in the second case, the previously mentioned structures were not observed, which corresponded to the optimized hydrodynamic sealing conditions.

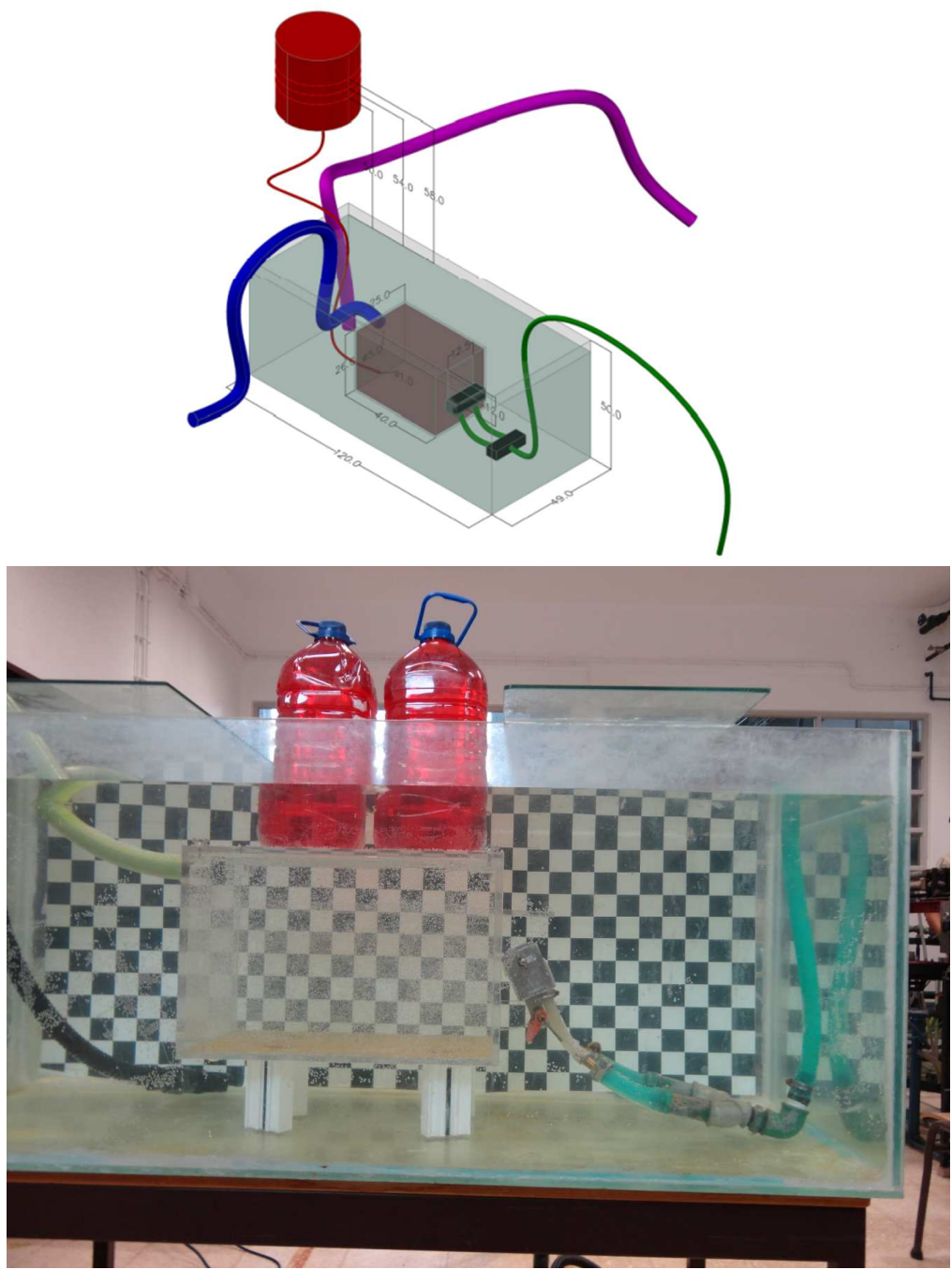

**Figure 3.** Test model—in the drawing, red corresponds to the contaminant circuit, dark blue to the exhaust circuit, green to the jet circuit, and purple to the water intake (dimensions in cm).

Table 1. Geometric characteristics of the jet.

| Test | Thickness of the Jet $b_0$ (m) | Width of the Jet L (m) | Section of the Jet (m²) | Height of the Door h (m) | Width of the Door w (m) |
|---|---|---|---|---|---|
| 57–88; 92–107 | 0.00125 | | 0.000182 | | |
| 89 | 0.00250 | 0.1456 | 0.000364 | 0.125 | 0.125 |
| 90 | 0.00375 | | 0.000546 | | |
| 91 | 0.00500 | | 0.000728 | | |

Furthermore, it was necessary to demonstrate that the water colorant intensity was strong enough to evidence any contaminant release through the curtain. Several tests were carried out in the same conditions, but by changing the water colorant concentration. In every test, the measurements were performed on the exhaust flow rates obtained for the optimum hydrodynamic sealing of the curtain. From test to test, the colorant concentration was increased until two different water colorant concentrations led to obtaining the same exhaust flow rates. From these two tests, the lower water colorant concentration was adopted in the subsequent research works.

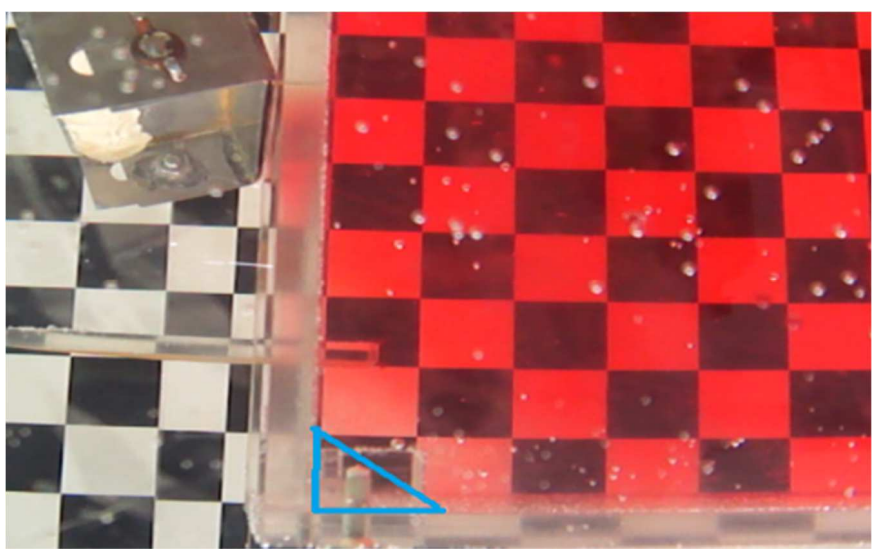

**Figure 4.** Oversized exhaust flow rate.

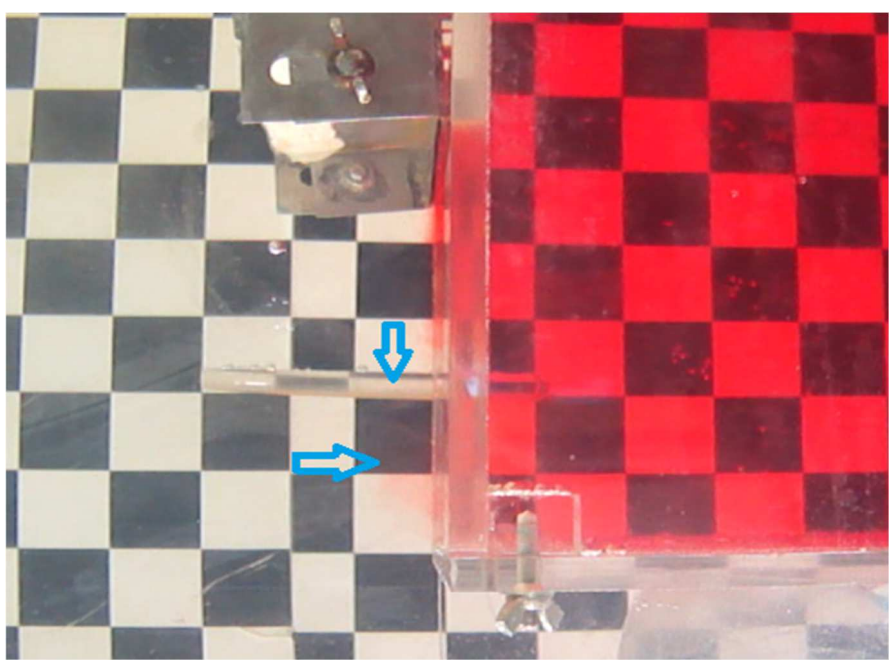

**Figure 5.** Insufficient exhaust flow rate.

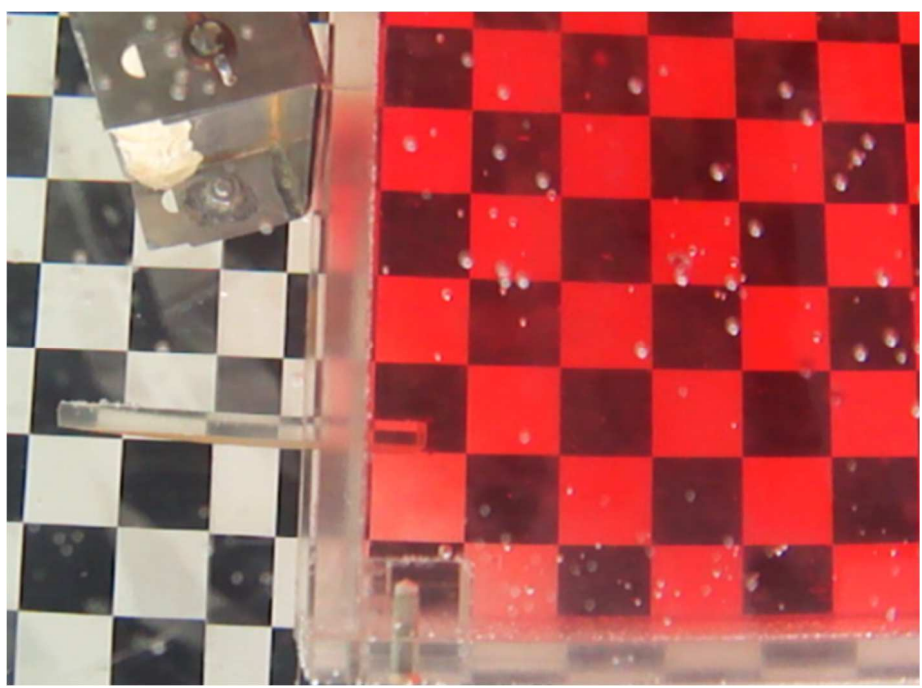

**Figure 6.** Limit state of the exhaust flow rate.

## 3. Results and Discussion

### 3.1. Test Results

The experimental results are presented in Tables 2 and 3. Every result corresponds to the average obtained from the two tests carried out in the same conditions. As the walls of the test compartment are impermeable and the water colorant supply flow rate is known, it is possible to assess the average velocity through the door $\overline{u_a}$, which is also presented in Table 2, measuring the exhaust flow rate $\dot{V}_{exaust}$ and computing the flow rate balance. The average velocity through the door $\overline{u_a}$ is put in evidence, because the authors believe that it is the relevant variable that influences the hydrodynamic sealing performance of the curtain.

**Table 2.** Test results obtained with both active (thickness of 0.00125 m) and inactive jets.

| Jet Thickness ($b_0$ = 0.00125 m) | | | | |
|---|---|---|---|---|
| $\overline{u_0}$ (m/s) | $\alpha_0$ (°) | $\dot{V}_{exaust}$ (L/s) | $\overline{u_a}$ (m/s) | Re |
| 1.70 | 0 | 1.20 | 0.077 | 2125 |
|  | 5 | 1.20 | 0.077 |  |
|  | 10 | 1.20 | 0.077 |  |
|  | 15 | 1.20 | 0.077 |  |
| 1.37 | 0 | 1.07 | 0.068 | 1710 |
|  | 5 | 1.07 | 0.068 |  |
|  | 10 | 1.07 | 0.069 |  |
|  | 15 | 1.07 | 0.069 |  |
| 0.98 | 0 | 0.84 | 0.054 | 1224 |
|  | 5 | 0.81 | 0.052 |  |
|  | 10 | 0.79 | 0.051 |  |
|  | 15 | 0.69 | 0.044 |  |
| 0.55 | 0 | 0.73 | 0.046 | 687 |
|  | 5 | 0.69 | 0.044 |  |
|  | 10 | 0.64 | 0.041 |  |
|  | 15 | 0.58 | 0.037 |  |
|  | 0 | 0.55 | 0.035 |  |
| 0.29 | 5 | 0.48 | 0.031 | 367 |
|  | 10 | 0.42 | 0.027 |  |
|  | 15 | 0.33 | 0.021 |  |
|  | 0 | 0.41 | 0.027 |  |
| 0.12 | 5 | 0.33 | 0.021 | 146 |
|  | 10 | 0.25 | 0.016 |  |
|  | 15 | 0.19 | 0.012 |  |
| 0.00 | - | 0.04 | 0.0026 | - |

**Table 3.** Records of experimental results with variations in jet thickness.

| $\overline{u_0}$ (m/s) | $\alpha_0$ (°) | $\dot{V}_{exaust}$ (L/s) | $\overline{u_a}$ (m/s) | Re | $b_0$ (m) |
|---|---|---|---|---|---|
| **0.29** |  | 0.33 | 0.021 |  | 0.00125 |
| **0.15** | 15 | 0.23 | 0.015 | 367 | 0.00250 |
| **0.10** |  | 0.20 | 0.013 |  | 0.00375 |
| **0.07** |  | 0.19 | 0.012 |  | 0.00500 |

*3.2. Analysis of the Influence of the Jet Nozzle Thickness on the Average Speed at the Door Opening*

The test results presented in Table 3 made it possible to study the influence of the jet nozzle thickness on the average velocity of the flow across the opening. Equation (19) was obtained by the linear regression, presented in Figure 7. It can be seen that the variation in the jet nozzle thickness, by keeping the Reynolds number constant (Re = 367), has no effect on the average velocity of the flow across the opening. It denotes that the average velocity of the flow across the opening depends linearly on the jet velocity at the nozzle only. Hence, it is not necessary to consider any influence associated with the jet nozzle thickness.

$$\overline{u_a} = 0.0414 \times \overline{u_0} - 0.0088 \text{ (m/s)} \tag{19}$$

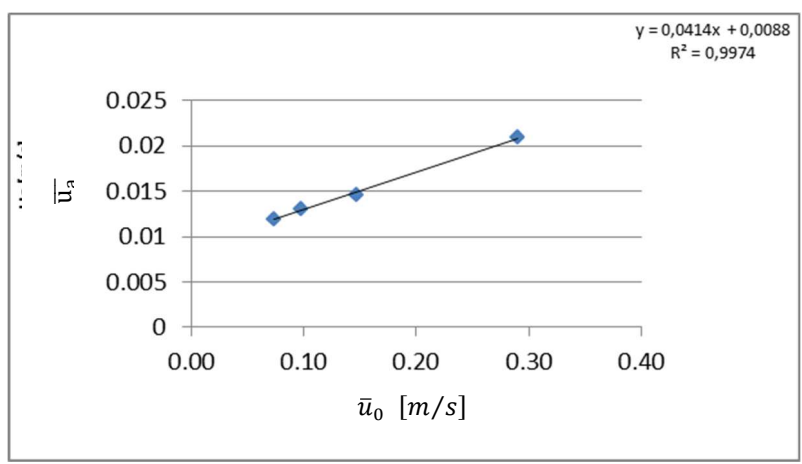

**Figure 7.** Analysis of the variation in the average velocity of the flow across the opening with jet velocity.

*3.3. Analysis of the Influence of the Jet Angle*

The second parameter analyzed was the influence of the jet angle. For the purpose of this study, the ratio between the average velocity at the opening and the nozzle velocity was compared with sin $\alpha$. The lines connecting the experimental results for the same Reynolds number are almost linear, if $(\overline{u}_a/\overline{u}_0)^{0.5}$ is used as a coordinate. Figure 8 shows that there is no influence of the jet angle for the Reynolds numbers close to Re = 1710 or above this value.

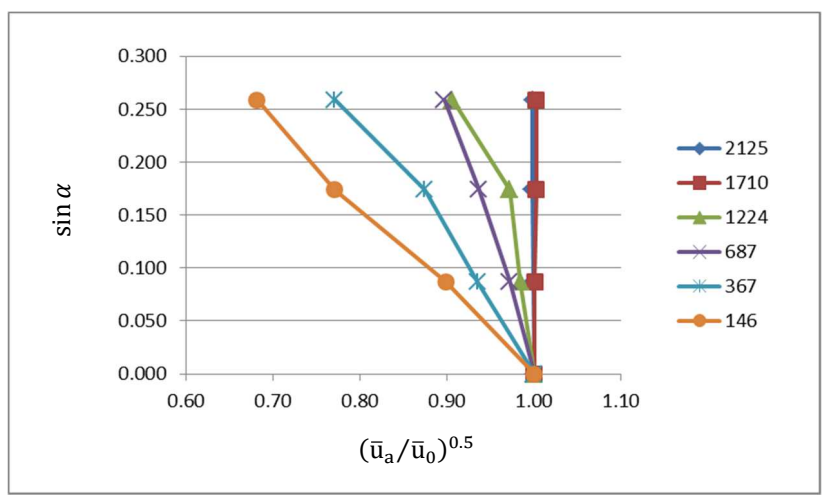

**Figure 8.** Jet angle influence.

### 3.4. Influence of $\overline{u_0}^{0.5} (1 - \sin \alpha)$ on the Average Speed at the Door Opening

When coordinates $\overline{u_a}$ and $\overline{u_0}^{0.5}(1 - \sin\alpha)$ are used to express the test results (Figure 9), it is possible to see that the test results almost collapse into a straight line. It should be noted that it was found that the angle does not influence the average velocity at the opening $\overline{u_a}$ for the Reynolds numbers of Re = 1710 and above, and therefore, it was deemed irrelevant to include the term $\sin \alpha$ in its expression (Equation (20)). Figure 10 shows the best fit for the test results when the coordinates referred to above are considered. The correlation coefficient obtained corresponded to 0.9883. Figure 9 also includes the test results obtained in the study about the effect of the variation on the jet thickness (see Table 3), designated in the figure key as 15*; it is clear that these results are fully aligned with further test results obtained for the same jet angle.

Equation (20), below, was deduced on the basis of the test results. Mention must be made of the fact that for Re ≥ 1710, the influence of $\sin \alpha$ was not introduced, as no influence of the angle has been included for that condition (as concluded before).

$$\overline{u_a} = \begin{cases} 0.0578 \times (1 - \sin\alpha) \times \sqrt{\overline{u_0}} + 0.0014 & \text{if Re} \leq 1224 \\ 0.0578 \times \sqrt{\overline{u_0}} + 0.0014 & \text{if Re} > 1224 \end{cases} \quad (20)$$

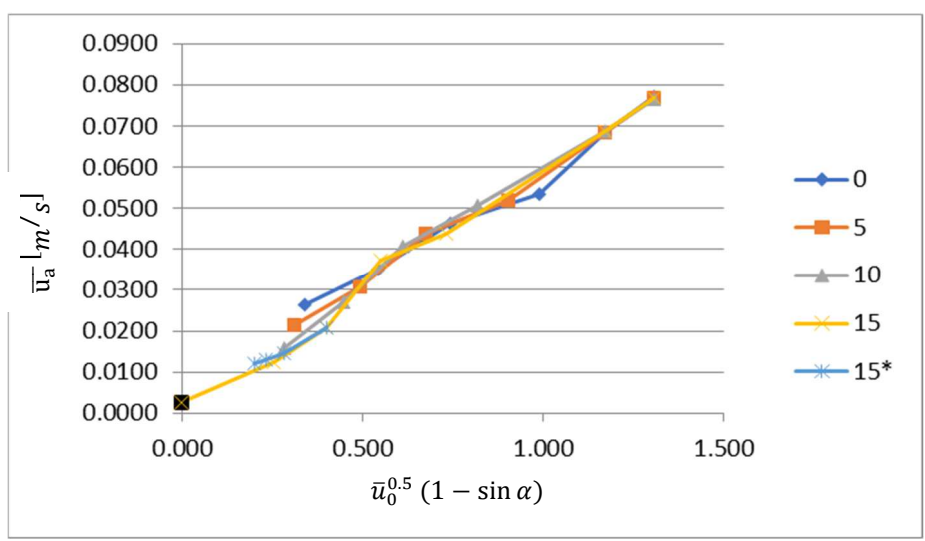

**Figure 9.** Relation between the average velocity at the opening and $\overline{u_0}^{0.5}(1 - \sin\alpha)$ for different jet angles (15* values correspond to tests with variation in jet thickness).

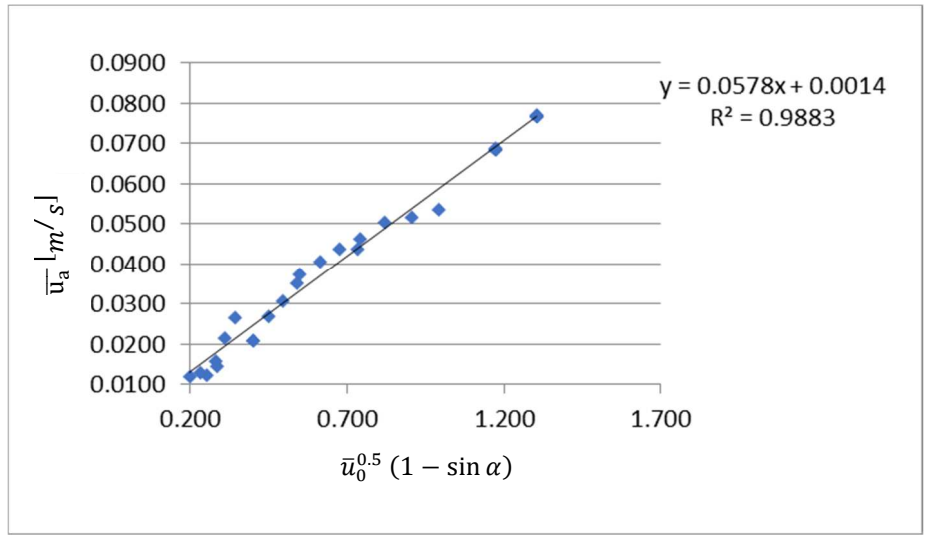

**Figure 10.** Linear regression between average velocity at the opening and $\overline{u_0}^{0.5} (1 - \sin \alpha)$.

Research was carried out on the best function that can be adjusted to the test results, having these coordinates as the basis. Therefore, the best fit using the least squares method was obtained for Equation (21). Variables A, B, C, f, and j were considered as the degrees of freedom. Table 4 presents the values for the variables corresponding to the best fit.

$$\overline{u_a} = A \times (B - \sin \alpha)^f \times \overline{u_0}^j + C \qquad (21)$$

**Table 4.** Results of the parameters adjusted by the least squares method.

| Variable | Value |
|---|---|
| A | 0.0595 |
| B | 1.00 |
| C | 0.00 |
| f | 1.05 |
| j | 0.470 |

As Figure 11 shows, a good adjustment of the experimental values has been achieved. This has led to obtaining a correlation coefficient of 0.9886 (Figure 12), which is quite close to the previous relation. These results are considered in Equation (22).

$$\overline{u_a} = \begin{cases} 0.0595 \times (1.00 - \sin \alpha)^{1.05} \times \overline{u_0}^{0.47} + 0.0002 & \text{if Re} \leq 1224 \\ 0.0595 \times \overline{u_0}^{0.47} + 0.0002 & \text{if Re} > 1224 \end{cases} \qquad (22)$$

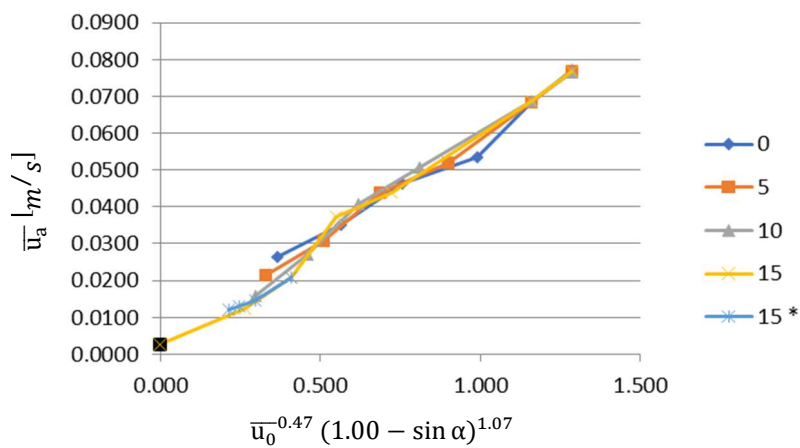

**Figure 11.** Relation between the average velocity at the opening and $\overline{u_0}^{0.47} (1.00 - \sin \alpha)^{1.07}$ (15* values correspond to tests with variations in jet thickness).

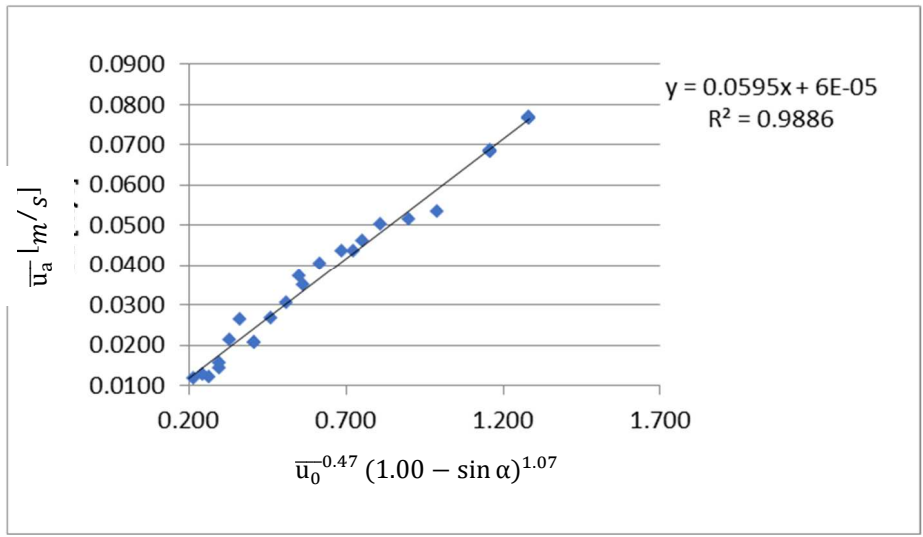

**Figure 12.** Linear regression between average velocity at the opening and $\overline{u_0}^{-0.47}(1.00 - \sin\alpha)^{1.07}$.

It was observed that some variables were very close to the unit or to 0.5. It was assumed that some deviation in relation to these numbers could be due to experimental errors and uncertainty. Therefore, the variables were rounded to B = 1.00, f = 0.50, and j = 1.00. Variable C was supposed to assume the value C = 0.0026 m/s, which is the velocity measured at the opening when the velocity of the jet is zero (see Table 2). Table 5 presents the values for the variables corresponding to the best fit, with A being the only degree of freedom.

**Table 5.** Rounded parameters and parameter A adjusted by the least squares method.

| Variable | Value |
| --- | --- |
| A | 0.0564 |
| B | 1.00 |
| C | 0.0026 |
| f | 1.00 |
| j | 0.50 |

As Figures 13 and 14 show, the test results are still approaching linearity, with a correlation coefficient of 0.987, when expressed by the appropriate variables.

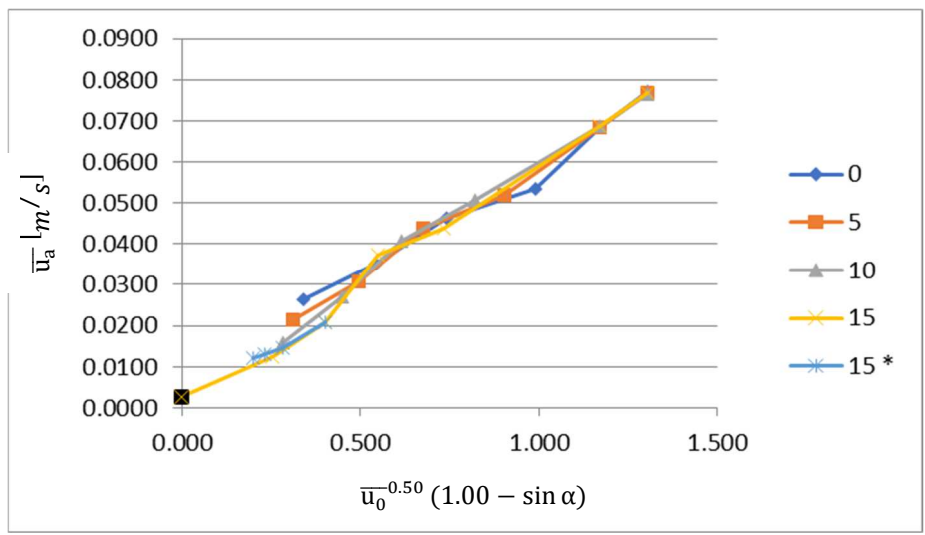

**Figure 13.** Relation between the average velocity at the opening and $\overline{u_0}^{-0.50}(1.00 - \sin\alpha)$ (15* values correspond to tests with variations in jet thickness).

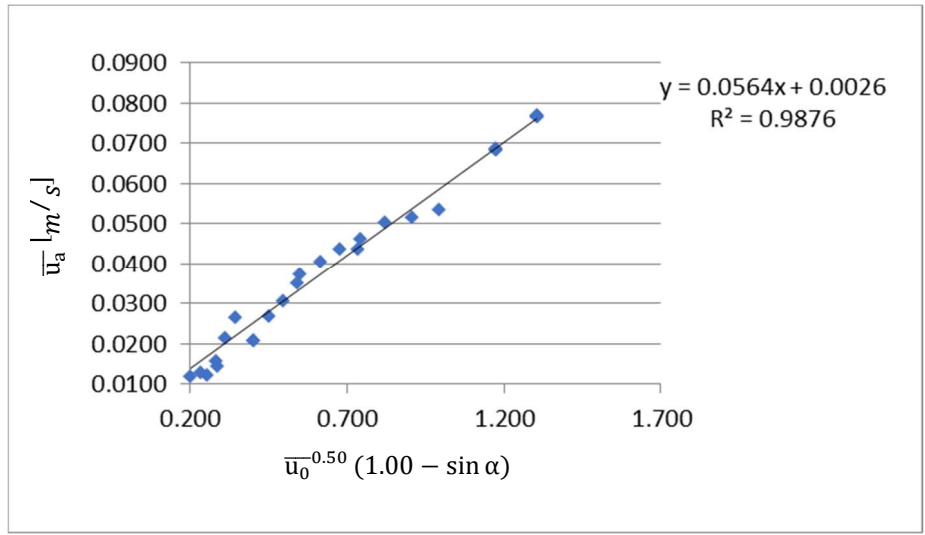

**Figure 14.** Linear regression average between the velocity at the opening and $\overline{u_0}^{-0.50}(1.00 - \sin\alpha)$.

As a result, the most appropriate equations to express the experimental results were obtained (Equations (23) and (24), in which the value $u_a = 0.0026$ was replaced, in a more general way, by $u_{a_{(u_0=0)}}$).

$$\overline{u_a} = \begin{cases} 0.0564 \times (1.00 - \sin\alpha) \times \overline{u_0}^{0.5} + 0.0026 & \text{if Re} \leq 1224 \\ 0.0564 \times \overline{u_0}^{0.5} + 0.0026 & \text{if Re} > 1224 \end{cases} \quad (23)$$

$$\overline{u_a} = \begin{cases} 0.0564 \times (1.00 - \sin\alpha) \times \overline{u_0}^{0.5} + u_{a_{(u_0=0)}} & \text{if Re} \leq 1224 \\ 0.0564 \times \overline{u_0}^{0.5} + u_{a_{(u_0=0)}} & \text{if Re} > 1224 \end{cases} \quad (24)$$

The graph in Figure 15 depicts the experimental results ($\overline{u_a}$; $\overline{u_0}$), which were compared with the corresponding values obtained by Equation (23). It is possible to observe that Equation (23) is a fair approximation of the experimental results. It is also clear that the results for the Reynolds number Re = 1224 (corresponding to $U_0$ = 0.98 m/s) are still in a transition between the low Reynolds number results, in which the jet angle is relevant, and the higher Reynolds number results, in which the jet angle is not relevant.

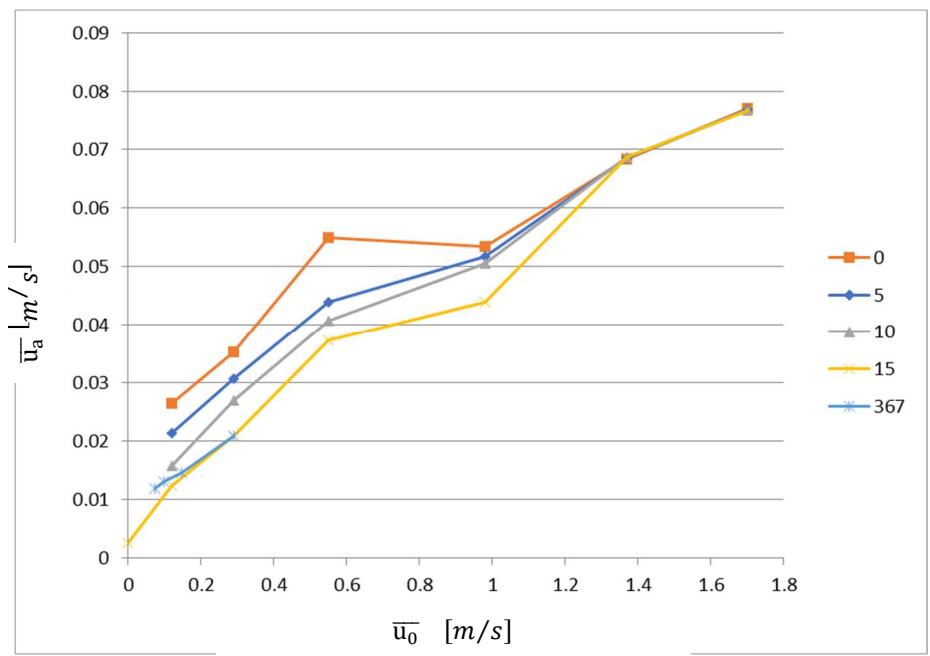

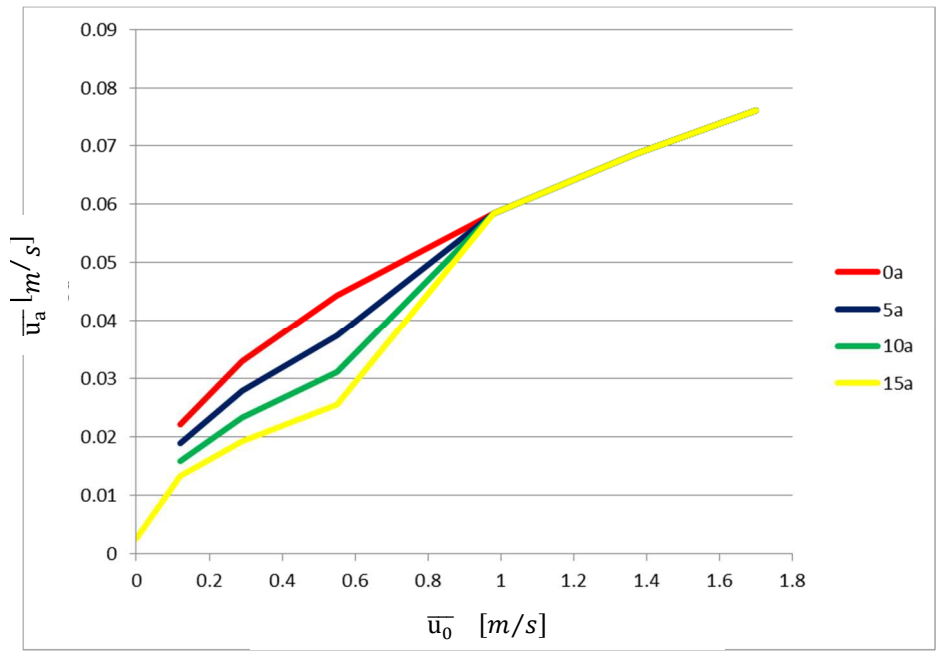

**Figure 15.** Comparison between experimental results and Equation (23).

*3.5. Analytical Method*

We stress that the empirical Function (23) does not show the same dependence on jet velocity and on jet angle as the theoretical simplified Function (14). Figure 16 presents Equation (14), together with the experimental results (the horizontal coordinate in this figure is different from the one adopted in the previous figures). It is clear that Equation (14) is close to the experimental results just for the jet angle of 15° and for Re < 1224. For the jet angles lower than 15°, the corresponding average jet nozzle velocity is higher than the predicted by Equation (14). For higher Reynolds numbers, Equation (14) predicts an average velocity of the flow across the door higher than the one measured in the experiments. This shows that some physical processes, which are relevant for the transport of the contaminant, are not expressed in the theoretically derived equation. We consider that such processes are related to the turbulent behaviour of the jet at the floor impingement zone.

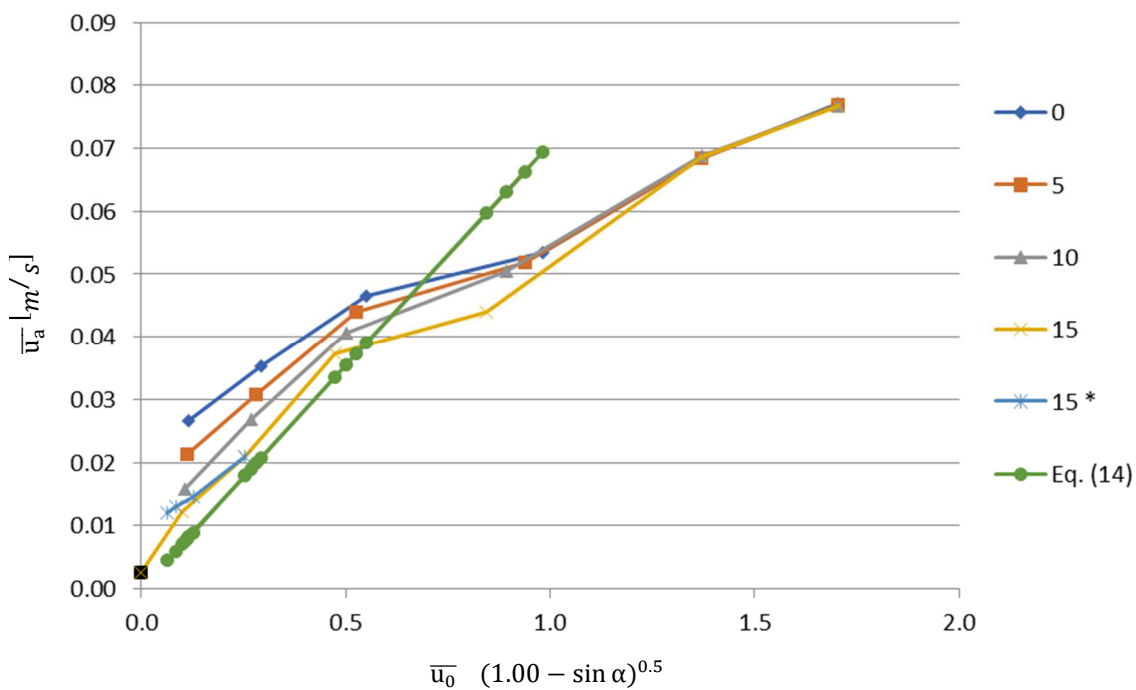

**Figure 16.** Comparison between experimental results and Equation (14).

Moreover, the experiments showed the following:

- The velocity at the opening is still necessary to avoid the spread of the contamination, when the jet is not active ($u_{a(u_0=0)}$);
- The jet angle exerts no influence if $Re \geq 1710$; and
- No influence of the nozzle thickness was observed in the experiments.

In a previous research project, using the same small-scale model, saltwater tests were carried out, with a view to test the use of a plane jet to avoid the dispersion of contaminant buoyancy driven through the plane jet [4]. In these tests, the nozzle velocity ranged from 0.142 m/s to 1.000 m/s (177< Re <3422), the nozzle thickness from 0.00125 to 0.00750 m, and the jet angle from 25° to 35°. Moreover, the parameters of the nozzle velocity, jet thickness, and jet angle were set before the beginning of the test, and the exhaust flow rate was adjusted during the test to achieve the hydrodynamic sealing of the curtain (the same methodology as the one followed in the tests reported in the previous section). In addition, the curtain velocity was reduced as much as possible, so as to find the lowest exhaust flow rate for every buoyant condition. In many tests, this optimal condition was not reached (the jet velocity was too high), and therefore, the turbulent mixing strongly influenced the hydrodynamic sealing of the curtain (as in the test results reported in the previous section). Although the saltwater test results are still strongly influenced by buoyancy, it is relevant to compare them with the experimental results reported in the previous section.

On basis of the saltwater tests, the average velocity of the flow across the door $\overline{u_a}$, which was necessary to avoid the contaminant transport through the plane jet, was co-related with the plane jet characteristics by Equation (25). Figure 17 compares the experimental values of $\overline{u_a}$ with the predictions of Equation (25). It shows that Equation (25) is able to predict fairly well the experimental results, but the dispersion of the test results is quite high because of the buoyancy effect. Equation (25) is similar to Equation (18), but multiplied by an empirical constant with a value of 1.178. As a result, Equation (25) indicates that the flow rate across the door must be approximately 17.8% above the flow rate driven by the jet from the non-contaminated compartment.

$$\overline{u_a} = 1.178 \times \left[ 0.22 \times \left( \frac{2 \times h}{b_0 \times \cos \alpha_0} \right)^{\frac{1}{2}} + 0.5 \right] \times \overline{u_0} \times \frac{b_0}{h} \qquad (25)$$

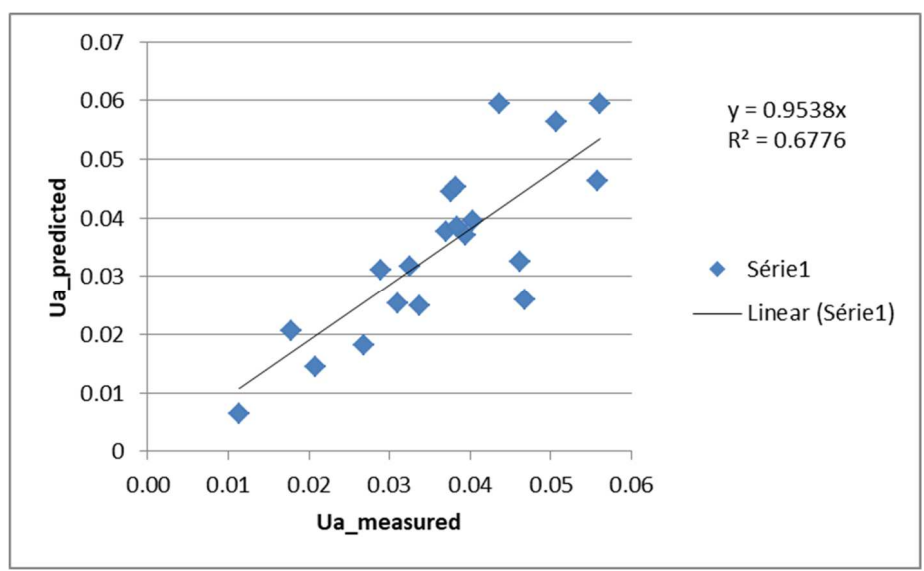

**Figure 17.** Comparison of the test results of $\overline{u_a}$ with the prediction of Equation (25).

Finally, Figure 18 compares the experimental results (left side) with Equation (23) (right side) and Equation (25). It is clear that for the Reynolds number Re > 1224 or for the jet angle of 15°, the experimental results show a similar development (but the dependence of $\overline{u_a}$ on $\overline{u_0}$ is not the same), even though the experimental results require a higher flow rate across the door to reach the hydrodynamic sealing of the curtain than one the predicted by Equation (25). It is relevant to notice that even in such different test conditions (with or without buoyancy), it is visible a similar trend.

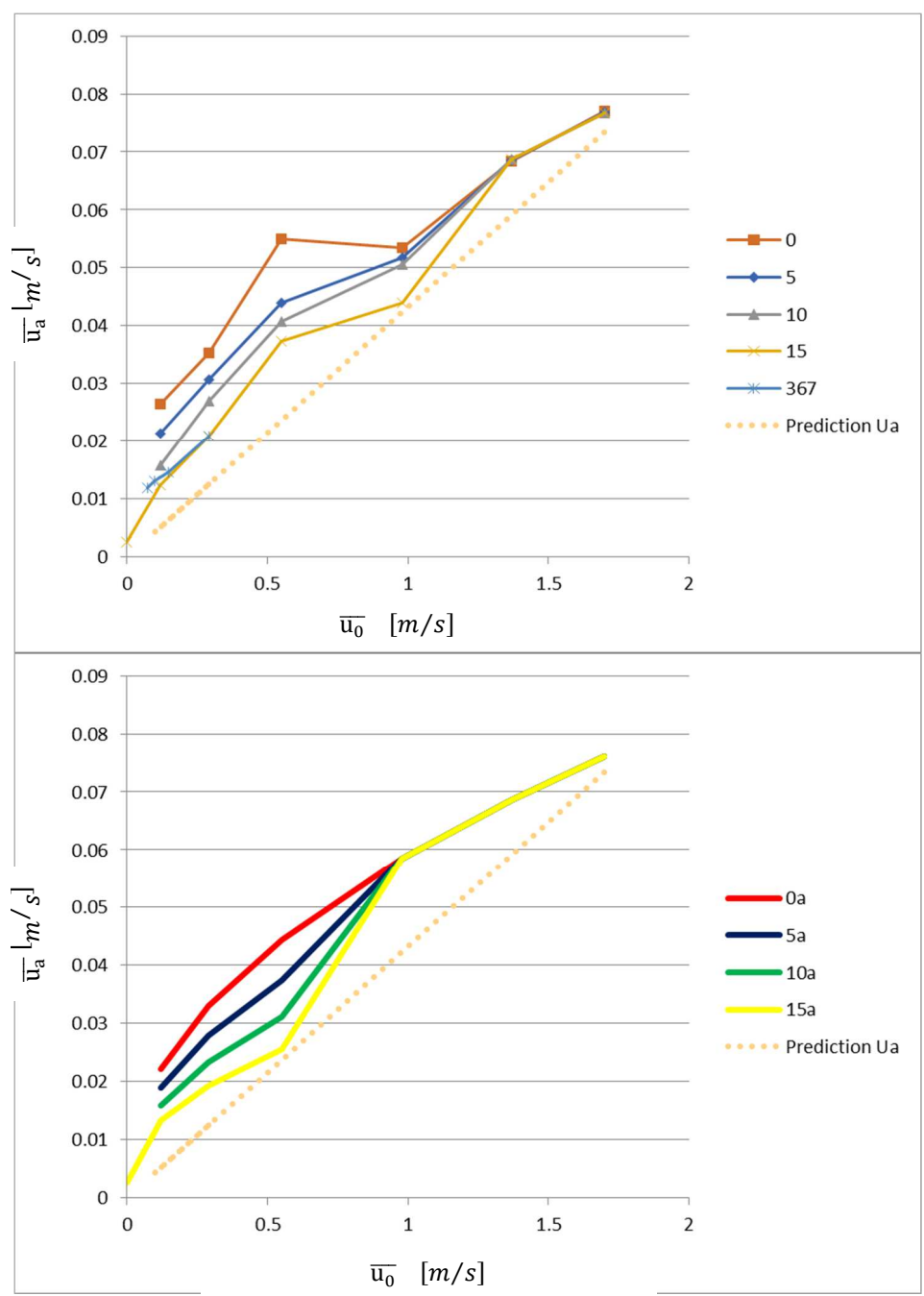

**Figure 18.** Comparison of Equation (25) (prediction Ua) with the test results of Section 3.1 and Equation (23).

## 4. Conclusions

In this research, small-scale water modelling was used to assess the hydrodynamic sealing of a curtain formed by a plane jet. This small-scale modelling is intended to reproduce the performance of a full-scale air curtain, with the Reynolds number being in the range between Re = 147 and Re = 2125. The authors believe that the empirical results obtained are innovative and representative of the optimal conditions to set the aerodynamic sealing of full-scale air curtains for the same Reynolds numbers in an isothermal environment.

The small-scale water modelling led to obtaining the empirical law presented in Equation (26), which corresponds to the limit state of the hydrodynamic sealing at the curtain.

$$\overline{u_a} = \begin{cases} 0.0564 \times (1.00 - \sin\alpha) \times \overline{u_0}^{0.5} + 0.0026 & \text{if Re} \leq 1224 \\ 0.0564 \times \overline{u_0}^{0.5} + 0.0026 & \text{if Re} > 1224 \end{cases} \quad (26)$$

Equation (26) shows that a simple model based on the flow momentum balance at the jet (Equation (14)) is unable to properly express the test results for higher Reynolds Numbers Re > 1224, but Equation (14) is close to the experimental results just for the jet angle of 15° and for low Reynolds Numbers (Re < 1224).

On the basis of previous small-scale saltwater test results [4], the average velocity of the flow across the door $\overline{u_a}$, which was necessary to avoid the contaminant transport through the plane jet, was co-related with the plane jet characteristics by Equation (27).

$$\overline{u_a} = 1.178 \times \left[ 0.22 \times \left( \frac{2 \times h}{b_0 \times \cos\alpha_0} \right)^{\frac{1}{2}} + 0.5 \right] \times \overline{u_0} \times \frac{b_0}{h} \quad (27)$$

For Reynolds number Re > 1224, the experimental results show a similar development (but the dependence of $\overline{u_a}$ on $\overline{u_0}$ is not the same), even though the experimental results require a higher flow rate across the door to reach the hydrodynamic sealing of the curtain than the one predicted by Equation (27).

**Author Contributions:** Investigation, F.O.; supervision, J.C.V. and D.A.

**Funding:** This project has received funding from the European Union's Horizon 2020 research and innovation programme, under the Marie Sklodowska-Curie grant agreement N 690968.

**Acknowledgments**: The small scale jet device was developed by Mr. Paulo Morais from Scientific Instrumentation Centre (CIC) of National Laboratory of Civil Engineering (LNEC). This device was built at CIC/LNEC.

**Conflicts of Interest:** The authors declare no conflict of interest. The funders had no role in the design of the study; in the collection, analyses, or interpretation of data; in the writing of the manuscript; and in the decision to publish the results.

**Nomenclature**

$b_0$—jet nozzle thickness
h—door height,
$J_1$—jet momentum
$J_2$—momentum of the flow rate rejected to the contaminated side
$J_3$—momentum of the flow rate rejected to the non-contaminated side
$J_a$—momentum of the flow through the door
L—jet width
$\dot{M}_0$—jet flow rate at the nozzle
$\dot{M}_1$—jet flow rate at the floor impingement zone
$\dot{M}_2$—flow rate rejected to the contaminated side
$\dot{M}_3$—flow rate rejected to the non-contaminated side
$Q_0$—flow rate at the jet nozzle
$Q_{jet}$—jet flow rate
$\overline{u_0}$ or U0—average jet nozzle velocity
$\overline{u_1}$—average jet velocity at the floor impingement zone
$\overline{u_2}$—average velocity of the flow rate rejected to the contaminated side
$\overline{u_3}$—average velocity of the flow rate rejected to the non-contaminated side
$\overline{u_a}$ or Ua—average velocity of the flow across the door
$u_{a(u_0=0)}$—average velocity of the flow across the door when $u_0 = 0$
$\dot{V}_{exaust}$—exhaust flow rate

    w—door width
    X—jet length, longitudinal coordinate of the jet
    $α_0$—slope of the jet
    ν—kinematic viscosity